\documentclass[twocolumn,floatfix,superscriptaddress,a4paper,showpacs,showkeys,nofootinbib,reprint,prl]{revtex4-1}
\usepackage{epsfig}
\usepackage{latexsym}
\usepackage{xspace}
\usepackage{hyperref}
\usepackage[latin2]{inputenc}
\usepackage{indentfirst}
\usepackage{enumerate}
\usepackage{color}

\usepackage[caption=false,position=top]{subfig}

\usepackage{amsmath}
\usepackage{amssymb}
\usepackage[english]{babel}
\usepackage{url}
\newcommand{\eq}[1]{\begin{align} #1 \end{align}}

\begin{document}


\title{
Cluster Expansion Model for QCD Baryon Number Fluctuations:\\
No Phase Transition at $\mu_B / T < \pi$
}

\author{Volodymyr Vovchenko}
\affiliation{
Institut f\"ur Theoretische Physik,
Goethe Universit\"at, Max-von-Laue-Str. 1, D-60438 Frankfurt am Main, Germany}
\affiliation{Frankfurt Institute for Advanced Studies, Giersch Science Center, Ruth-Moufang-Str. 1, D-60438 Frankfurt am Main, Germany}

\author{Jan Steinheimer}
\affiliation{Frankfurt Institute for Advanced Studies, Giersch Science Center, Ruth-Moufang-Str. 1, D-60438 Frankfurt am Main, Germany}

\author{Owe Philipsen}
\affiliation{
Institut f\"ur Theoretische Physik,
Goethe Universit\"at, Max-von-Laue-Str. 1, D-60438 Frankfurt am Main, Germany}

\author{Horst Stoecker}
\affiliation{
Institut f\"ur Theoretische Physik,
Goethe Universit\"at, Max-von-Laue-Str. 1, D-60438 Frankfurt am Main, Germany}
\affiliation{Frankfurt Institute for Advanced Studies, Giersch Science Center, Ruth-Moufang-Str. 1, D-60438 Frankfurt am Main, Germany}
\affiliation{
GSI Helmholtzzentrum f\"ur Schwerionenforschung GmbH, Planckstr. 1, D-64291 Darmstadt, Germany}

\begin{abstract}
A Cluster Expansion Model (CEM), representing a relativistic extension of Mayer's cluster expansion, 
is constructed to study baryon number fluctuations in QCD.
The temperature dependent first two coefficients,
corresponding to the partial
pressures in the 
baryon number $B = 1$ and $B = 2$ sectors, are the only model input, which we fix by recent lattice data
at imaginary baryochemical potential.
All other coefficients are constructed in terms of the first two and required to match the 
Stefan-Boltzmann limit 
of non-interacting quarks and gluons 
at $T \to \infty$.
The CEM allows calculations 
of the baryon number susceptibilities $\chi_k^B$  
to arbitrary order.
We obtain excellent agreement with 
all
available lattice data for the baryon number fluctuation measures $\chi_2^B$, $\chi_4^B$, $\chi_6^B$ 
and predict higher order susceptibilities, that are not yet available from Lattice QCD.
The calculated susceptibilities are then used to extract the radius of convergence of the Taylor expansion of the pressure. 
The commonly used ratio test fails due to the non-trivial asymptotic behavior of the Taylor coefficients. 
At the same time, a more elaborate estimator provides finite convergence radii at all  
considered
temperatures and in  
agreement with the singularities of Pad\'e approximants.
The associated singularities lie in the complex $\mu_B/T$-plane and appear smoothly connected to the Roberge-Weiss transition at high temperatures and imaginary chemical potential.
No evidence for a phase transition at $\mu_B / T \lesssim \pi$ 
and $T>135$~MeV
is found.

\end{abstract}



\maketitle


The grand canonical thermodynamic potential of QCD is
an even function of the real baryon chemical potential 
$\mu_B$ because of the CP-symmetry.
Taking the quantization of the baryon charge into account, the 
QCD pressure
has a fugacity expansion 
\eq{
p/T^4 = \frac{1}{VT^3} \ln Z = \sum_{k = 0}^{\infty} \, p_k (T) \, \cosh\left( \frac{k \, \mu_B}{T} \right).
}
This relation can formally be viewed as a relativistic extension of Mayer's cluster expansion in fugacities~\cite{GNS}.
A similar formalism is also employed in the canonical approach~\cite{Hasenfratz:1991ax,Alexandru:2005ix,deForcrand:2006ec,Nagata:2010xi}, where the fugacity expansion is applied to the partition function $Z$ itself~(see Refs.~\cite{Nakamura:2015jra,Bornyakov:2016wld} for recent developments).

The 
net baryon density reads
\eq{\label{eq:chi1real}
\frac{\rho_B}{T^3} 
= \frac{\partial (p/T^4)}{\partial (\mu_B / T)}
= \sum_{k = 1}^{\infty} \, b_k (T) \, \sinh \left( \frac{k \, \mu_B}{T} \right),
}
where
$b_k (T) \equiv k \, p_k(T)$.
Analytic continuation to an imaginary chemical potential yields a purely imaginary $\rho_B / T^3$, with $b_k(T)$ becoming its Fourier expansion coefficients.
The analytic continuation to/from imaginary $\mu$ is one of the methods used to study QCD
at finite net baryon density~\cite{deForcrand:2002hgr,DElia:2002tig,deForcrand:2003vyj,DElia:2004ani,deForcrand:2010ys,Bonati:2014kpa,Cea:2015cya,Bluhm:2007cp,Morita:2011jva}.
The leading four Fourier coefficients $b_k$ were studied in Refs.~\cite{BorsanyiQM2017,Vovchenko:2017xad} using lattice simulations 
with physical quark masses at imaginary $\mu_B$, as well as  phenomenological models.

At low temperatures, below the crossover transition, 
a behavior similar to an uncorrelated gas of hadrons, i.e.~$| b_k(T) / b_1(T) | \ll 1$ for $k \geq 2$, is seen in lattice QCD~(LQCD)~\cite{BorsanyiQM2017,Vovchenko:2017xad}.
Lattice simulations at $T > 135$~MeV yield negative $b_2(T)$ values, which could be interpreted in terms of repulsive baryonic interactions.
The first four LQCD Fourier coefficients, up to $T \simeq 185$~MeV, can indeed be described rather well by a hadron resonance gas (HRG) model with excluded-volume (EV) interactions between the (anti)baryons, with one canonical ``eigenvolume'' parameter $b \simeq 1$~fm$^3$ for all (anti)baryon species~\cite{Vovchenko:2017xad}.

QCD in the high-temperature limit resembles an ideal gas of massless quarks and gluons. 
In this Stefan-Boltzmann (SB) limit, 
the quarks carry a fractional baryon charge of $1/3$, and
the coefficients $b_k$ read~\cite{Vovchenko:2017xad}
\eq{\label{eq:SB}
b_k^{\rm {_{SB}}} = \frac{(-1)^{k+1}}{k} \, \frac{4 \, [3 + 4 \, (\pi k)^2]}{27 \, (\pi k)^2}.
}

The fugacity series~\eqref{eq:chi1real} is a useful tool because it treats the non-interacting hadron limit at low temperatures and the non-interacting quark limit at high temperatures in the same framework.
In this work a model is constructed which allows to calculate the coefficients $b_k$ at all intermediate temperatures between these two limiting cases.
This Cluster Expansion Model (CEM) is based on the following assumptions:
\begin{itemize}
\item 
The first coefficient $b_1(T)$ -- the QCD partial pressure in the $|B| = 1$ sector -- is taken as input.
It is interpreted as a temperature dependent density of ``free'' excitations with $B = \pm 1$.
\item
The second coefficient, $b_2(T)$, is also taken as input.
In the spirit of a cluster expansion it parametrizes 
the baryon-baryon interactions.
In the HRG-EV model $b_2$ is rewritten as
\eq{\label{eq:b2}
b_2 (T) & = -b(T) \, T^3 \, [b_1(T)]^2,
}
where $b(T)$ is a temperature dependent ``coupling'' parameter.
\item
 Mayer's cluster expansion assumes two-baryon interactions only, 
expected to be a good approximation at sufficiently low density or high temperature,
i.e.~moderate $\mu_B/T$. 
The higher-order coefficients $b_k(T)$ are then 
expressed in terms of the first two, motivated by a
HRG-EV-type system with two-particle hard core interactions \cite{Vovchenko:2017xad}:
\eq{\label{eq:bk}
b_k (T) & = \alpha_k \, [-b(T) \, T^3]^{k-1} \, [b_1(T)]^k 
\nonumber
\\
& =
\alpha_k \, \frac{[b_2(T)]^{k-1}}{[b_1(T)]^{k-2}},
}
the $\alpha_k$ are temperature \emph{independent} parameters.
\item
The model is constrained by the SB limit~\eqref{eq:SB} of massless quarks and gluons at high temperatures\footnote{This SB limit constraint is an important new element compared to an earlier study in Ref.~\cite{Vovchenko:2017xad}.}, i.e. $\displaystyle b_k(T) \to b_k^{\rm {_{SB}}}$ as $T \to \infty$.
Assuming $b_1(T) \to b_1^{\rm {_{SB}}}$ and $b_2(T) \to b_2^{\rm {_{SB}}}$, this condition fixes the coefficients $\alpha_k$:
\eq{\label{eq:alphak}
\alpha_k = 
\frac{[b_1^{\rm {_{SB}}}]^{k-2}}{[b_2^{\rm {_{SB}}}]^{k-1}} \, b_k^{\rm {_{SB}}}.
}

\end{itemize}
Eqs.~\eqref{eq:SB}-\eqref{eq:alphak} define \emph{all} coefficients $b_k(T)$ in CEM, using only $b_1(T)$ and $b_2(T)$ as input.

In what we term CEM-LQCD, $b_1(T)$ and $b_2(T)$ are
fixed by recent (2+1)-flavor, $N_{\tau} = 12$ lattice QCD simulations at imaginary $\mu_B$ of the Wuppertal-Budapest collaboration~\cite{Vovchenko:2017xad}.
In an alternative CEM-HRG,
$b_1(T)$ and $b_2(T)$ are taken from the HRG-EV model with a constant
$b(T) = 1$~fm$^3$ value~\cite{Huovinen:2017ogf,Vovchenko:2017xad,Vovchenko:2017drx}.

Note that, for a calculation of the pressure using the CEM, also the partial pressure $p_0(T)$ in the $|B| = 0$ sector is required as input.
Here we only study baryon number fluctuations for which this is not needed.

\begin{figure}[t]
  \centering 
  \includegraphics[width=.49\textwidth]{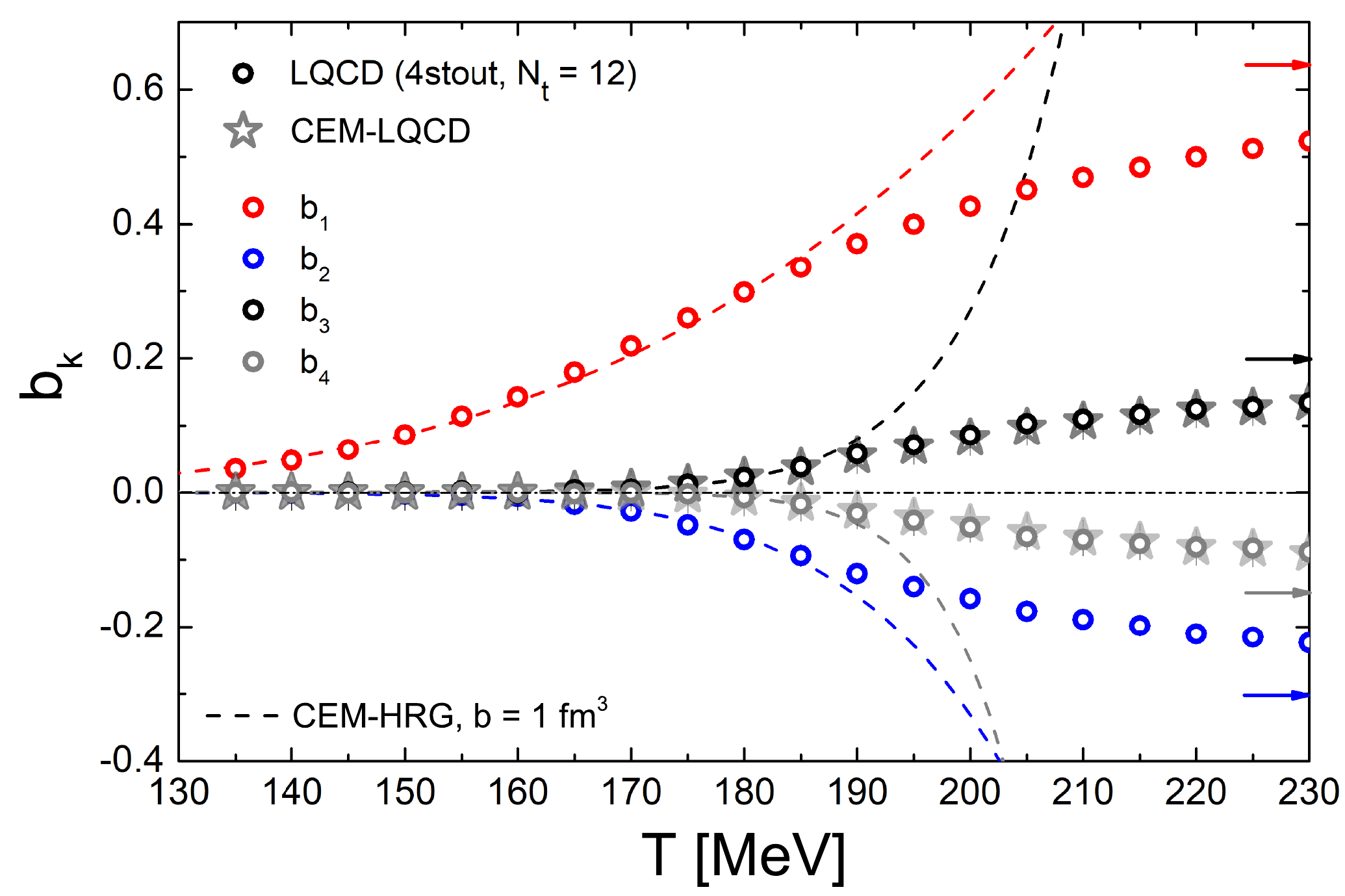}
  \caption{The temperature dependence of the first four Fourier coefficients $b_k$~\eqref{eq:chi1real}. 
  Lattice QCD results from imaginary $\mu_B$ simulations~\cite{Vovchenko:2017xad} are depicted by the circles,
  the calculations of $b_3$ and $b_4$ within the CEM-LQCD 
  are depicted by the stars.
  Predictions of the CEM-HRG model with $b(T) = 1$~fm$^3$ are depicted by the dashed lines.
  The arrows correspond to the Stefan-Boltzmann limit~\eqref{eq:SB}.
  } 
  \label{fig:bk}
\end{figure}

\begin{figure*}[t]
\begin{center}
\includegraphics[width=\textwidth]{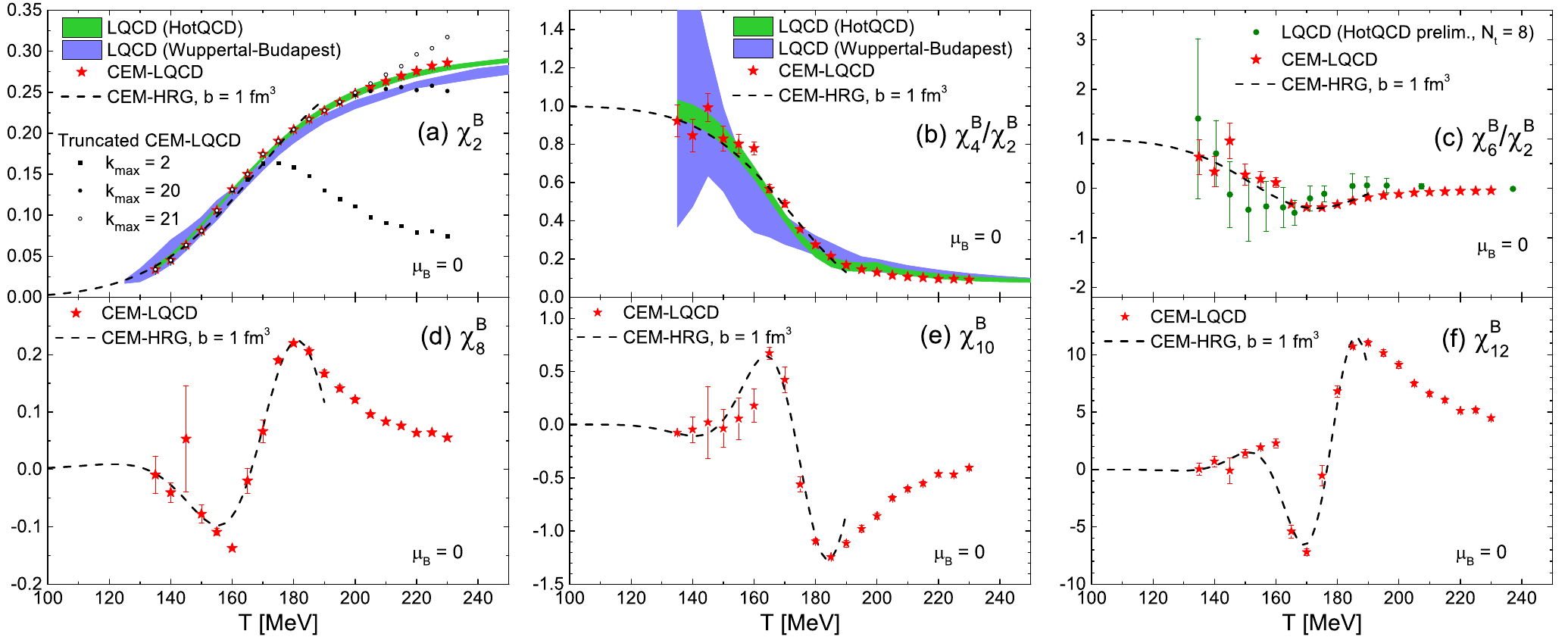}
\caption{The temperature dependences of the net baryon number susceptibilities at $\mu_B = 0$.
These include (a) $\chi_2^B$,
(b) $\chi_4^B / \chi_2^B$,
(c) $\chi_6^B / \chi_2^B$,
(d) $\chi_8^B$,
(e) $\chi_{10}^B$,
and (f) $\chi_{12}^B$.
Calculations within CEM-LQCD 
and CEM-HRG~($b = 1$~fm$^3$) models are depicted by the red stars and dashed black lines, respectively.
Lattice QCD results of the Wuppertal-Budapest~\cite{Borsanyi:2011sw,Bellwied:2015lba} and HotQCD~\cite{Bazavov:2017dus,Bazavov:2017tot} collaborations are shown, respectively, by the blue and green bands where available.
}
\label{fig:chin}
\end{center}
\end{figure*}

Temperature dependences of the first four coefficients $b_k(T)$, as calculated in lattice QCD simulations~\cite{Vovchenko:2017xad}, the CEM-LQCD model, and the CEM-HRG model, are shown in Fig.~\ref{fig:bk} by the circles, the stars, and the dashed lines, respectively.
The CEM-LQCD parametrization reproduces
the lattice data for $b_1$ and $b_2$ by construction. However, both
$b_3$ and $b_4$ are \emph{predicted} by the CEM-LQCD model~[Eq.~\eqref{eq:bk}]
and they agree quantitatively 
with the lattice data for all temperatures $135 \leq T \leq 230$~MeV.
The validity of Eq.~\eqref{eq:bk} for higher-order coefficients can be checked by future lattice simulations at imaginary $\mu_B$.

The CEM-HRG model reproduces the same coefficients up to $T \simeq 185-190$~MeV, however, it rapidly diverges from the lattice data 
at higher temperatures.

The criterion for the convergence 
of the expansion in Eq.~~\eqref{eq:chi1real} within CEM
is given by the ratio test:
\eq{\label{eq:converge}
\lim_{k \to \infty} \left| \frac{b_{k+1}(T) \sinh\left[\frac{(k+1) \, \mu_B}{T}\right]}{b_{k}(T) \sinh\left[\frac{k \, \mu_B}{T}\right]} \right| = \left| \frac{b_2(T) \, b_1^{\rm {_{SB}}} }{b_1(T) \, b_2^{\rm {_{SB}}} } \right| e^{ \frac{|\mu_B|}{T} }  < 1.
}
This condition is violated at sufficiently high $\mu_B / T$ values.
At $\mu_B = 0$, the condition is satisfied at all temperatures considered for the CEM-LQCD.
In the CEM-HRG at $\mu_B = 0$, it is satisfied only up to $T \simeq 195$~MeV,
which
correlates with the breakdown of CEM-HRG 
shown in Fig.~\ref{fig:bk}.

In the following, the net baryon number susceptibilities $\chi_{2n}^B(T)$ are calculated at zero baryo\-chemical potential. They are defined as follows:
\eq{\label{eq:chikdef}
\chi_{2n}^B(T) \equiv \left. \frac{\partial^{2n} (p/T^4)}{\partial (\mu_B / T)^{2n}} \right|_{\mu_B = 0} = \sum_{k = 1}^{\infty} \, k^{2n - 1} \, b_k(T).
}
The $\chi_{2n}^B$ are proportional to the coefficients of the Taylor expansion of the QCD pressure with respect to $\mu_B / T$. 
The odd-order susceptibilities, $\chi_{2n-1}^B(T)$, all vanish at $\mu_B = 0$.

The temperature dependence of the net baryon number susceptibilities, up to $\chi_{12}^B$, 
is shown in Fig.~\ref{fig:chin} by red stars~(CEM-LQCD) and dashed black lines~(CEM-HRG), respectively.
The error bars shown for the CEM-LQCD calculations 
were obtained by standard statistical propagation of
the uncertainties in the input data,
i.e.~by calculating the derivatives of the observables with respect to $b_1(T)$ and $b_2(T)$. We cross checked the validity 
by varying
$b_1(T)$ and $b_2(T)$ within their uncertainties and observing 
consistent
variations in the observables.

In our calculations, the sum~\eqref{eq:chikdef} is truncated at a sufficiently high value $k = k_{\rm max}$, such that the terms with $k > k_{\rm max}$ have negligible contributions to $\chi_{2n}^B$ at a given temperature $T$.
Fig.~\ref{fig:chin}(a) shows that the first two terms~($k_{\rm max} = 2$, i.e.~the lattice input), reproduce the full result for $\chi_2^B$ only at moderate temperatures, $T \lesssim 160$~MeV.
Hence, a higher number of terms is required to correctly calculate the susceptibilities at higher temperatures: For $k_{\rm max} = 20$, the full result for $\chi_2^B$ is reproduced up to $T \simeq 200$~MeV,
while $k_{\rm max} \simeq 80$ is required to calculate $\chi_2^B$ at $T = 230$~MeV. 
In order to cope with the large round-off errors, which arise in the numerical calculations, 
the arbitrary precision arithmetic provided by the \emph{Mathematica} package is employed.

Lattice QCD results for $\chi_2^B$, $\chi_4^B / \chi_2^B$, and $\chi_6^B / \chi_2^B$ of the Wuppertal-Budapest~\cite{Borsanyi:2011sw,Bellwied:2015lba} and HotQCD~\cite{Bazavov:2017dus,Bazavov:2017tot} collaborations are also shown in Fig.~\ref{fig:chin}, where available.
Comparing the CEM results with the lattice data elucidates the excellent predictive power of the present CEM-LQCD approach.
A precise calculation of the $\chi_{2n}^B$ values 
requires summation over tens and, in some cases, hundreds of $b_k(T)$ coefficients. 
All of them, except the first two, are \emph{predicted} by CEM.
The CEM-LQCD \emph{predictions} for $\chi_8^B$, $\chi_{10}^B$, and $\chi_{12}^B$ are shown in Fig.~\ref{fig:chin}(d)-(f).
The comparison with future lattice data will be able to confirm~(or refute) the validity of the CEM approach presented here.

The CEM-HRG model results, as shown by the dashed lines in Fig.~\ref{fig:chin}, agree very well with CEM-LQCD calculations, up to $T \simeq 180$~MeV, for \emph{all} considered observables.
Hence, the drastic temperature dependence of the baryon number fluctuations in this temperature range, as well as the particularly strong deviations from the ideal HRG baseline -- the Skellam distribution --  are convincingly interpreted in terms of repulsive baryonic interactions~(see also~\cite{Vovchenko:2016rkn,Huovinen:2017ogf,Vovchenko:2017xad}).

\begin{figure}[t]
  \centering 
  \includegraphics[width=.49\textwidth]{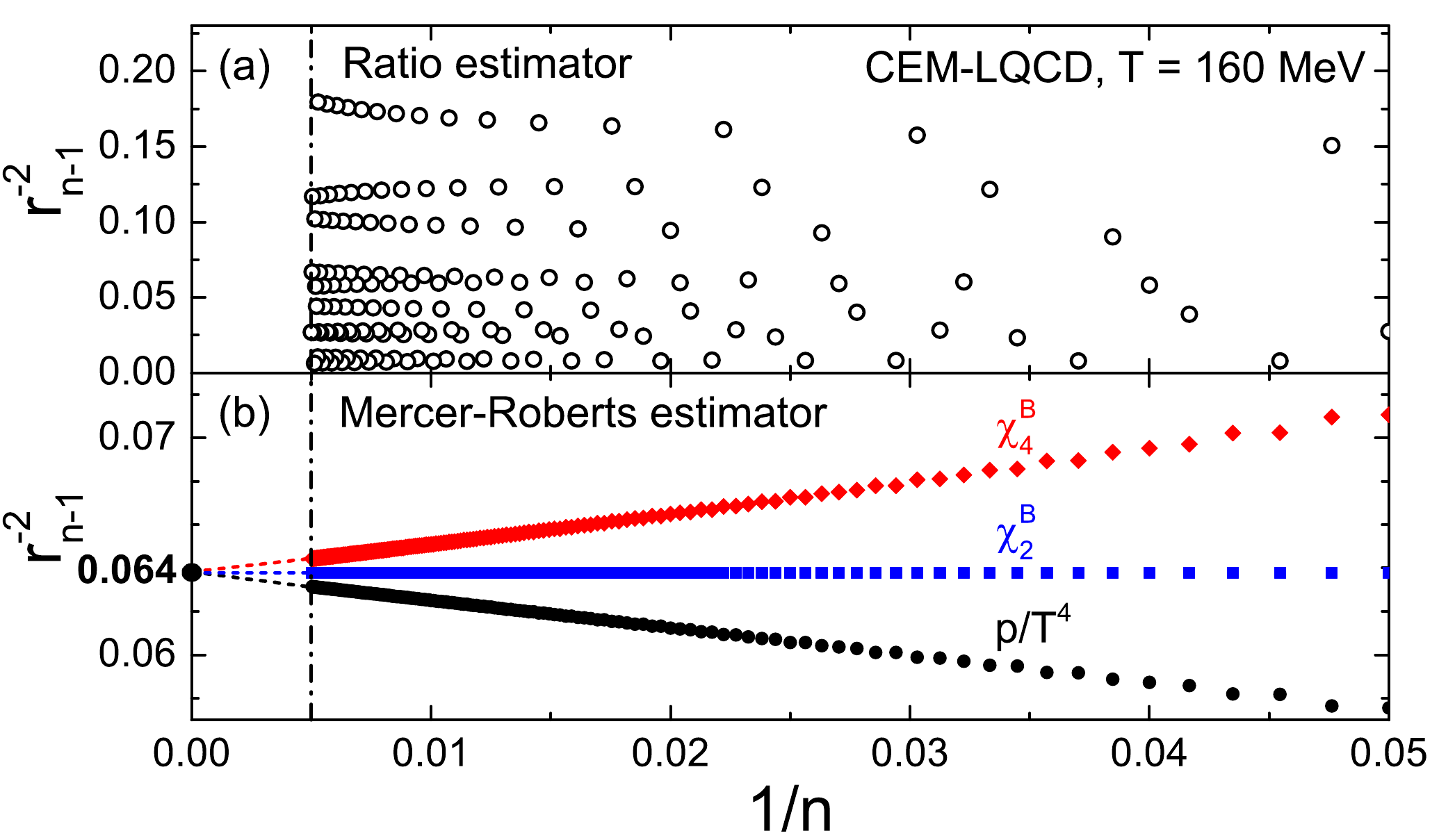}
  \caption{The Domb-Sykes $1/r_{n-1}^2$ vs $1/n$ plots, calculated within the CEM-LQCD model at $T = 160$~MeV using (a) ratio~(open symbols) and (b) Mercer-Roberts~(full symbols) estimators for radius of convergence of the Taylor expansion of $p/T^4$~(circles), $\chi_2^B$~(squares), and $\chi_4^B$~(diamonds).
The linear extrapolations of the Mercer-Roberts estimators to $1 / n = 0$ are depicted by the dashed lines ending at a circle. 
  } 
  \label{fig:DS}
\end{figure}

The ability of the CEM-formalism to calculate baryon number susceptibilities to very high order 
provides a unique opportunity to analyze the radius of convergence of the Taylor expansion of the QCD pressure,
\eq{\label{eq:Taylor}
\frac{p(T,\mu_B)-p(T,0)}{T^4} =
\sum_{n = 1}^{\infty} \, \frac{\chi_{2n}(T)}{(2n)!} \, \left( \frac{\mu_B}{T} \right)^{2n}~.
}

The radius of convergence, $r_{\mu/T}$, of this series 
at a given temperature corresponds to the distance to the nearest singularity in 
the complex $\mu_B / T$ plane and this has been used in various 
attempts to constrain the location of the critical point of QCD by numerical evaluation of a few leading 
coefficients in lattice QCD~\cite{Allton:2002zi,Gavai:2004sd,Allton:2005gk} or in effective models~\cite{Stephanov:2006dn,Karsch:2010hm,Skokov:2010uc}.
Derivatives of the pressure series expansion may be used equally well.
In the present work, 
estimates based on the Taylor series of $p/T^4$, $\chi_2^B$, and $\chi_4^B$ are analyzed.

First the ratio estimator, $r_n = \left| c_n / c_{n+1}\right|^{1/2}$, is used. 
The square root in this estimator~[as well as the extra square root in Eq.~\eqref{eq:MR}] appears due to the fact that the Taylor expansion~\eqref{eq:Taylor} is actually in $(\mu_B/T)^2$ rather than just in $\mu_B/T$.
Here $c_n = \chi_{2n} / (2n)!$ for the $p/T^4$ expansion,
$c_n = \chi_{2n} / (2n-2)!$ for the $\chi_2^B$ expansion, and
$c_n = \chi_{2n} / (2n-4)!$ for the $\chi_4^B$ expansion.
The $n \to \infty$ limit of $r_n$, if it exists, is the same for all three expansions and corresponds to the true radius of convergence.
This limit can be determined with the Domb-Sykes presentation~\cite{DombSykes}, by plotting $1/r_{n-1}^2$ versus $1/n$ for a finite number of terms, and then  extrapolating the result linearly to $1 / n = 0$.
To illustrate the behavior of the $r_{\mu/T}$ estimators we show 
$T = 160$~MeV as an example, the behavior at all other temperatures 
investigated is similar.
The Domb-Sykes plot for the Taylor series of $p/T^4$, as obtained within the CEM-LQCD model at $T = 160$~MeV by using the first 200 terms of the Taylor expansion, is depicted in Fig.~\ref{fig:DS}a by the open symbols.
(The plots for $\chi_2^B$ and $\chi_4^B$ are similar and not shown).
Note how the different orders jump between several branches of $1/r_{n-1}^2$ as $1/n$ approaches zero, with no unique limiting value in sight.
This behavior is caused by the irregular asymptotic structure of the Taylor coefficients.
Convergence of a Domb-Sykes plot for the ratio test requires the coefficients to asymptotically be of the same sign or to alternate in sign.
Neither of the two scenarios is realized in the CEM-LQCD: even at very high order, at least two positive- and at least two negative coefficients appear regularly in a row.
Therefore, 
the ratio estimator does not give a correct estimate of 
$r_{\mu/T}$
since the limit $\displaystyle \lim_{n \to \infty} r_n$ simply does not exist.
(Note that the ratio estimator is commonly used in the lattice QCD studies of the Taylor expansion~\cite{DElia:2016jqh,Datta:2016ukp,Bazavov:2017dus}).

More elaborate estimators do exist which deal with the irregular asymptotic structure of the Taylor coefficients.
Consider the Mercer-Roberts estimator~\cite{MercerRoberts},
\eq{\label{eq:MR}
r_n = \left| \frac{c_{n+1} \, c_{n-1} - c_n^2 } { c_{n+2} \, c_n - c_{n+1}^2 }\right|^{1/4}.
}
The corresponding $1/r_{n-1}^2$ vs $1/n$ plot is shown by the full symbols in Fig.~\ref{fig:DS}b.
For all three Taylor expansions, the Mercer-Roberts estimators appear to converge to the same point
as
$1 / n \to 0$.
Linear extrapolations to $1 / n \to 0$ give a value for the radius of convergence $r_{\mu/T}$.
The behavior of both estimators shown in Fig.~\ref{fig:DS} is similar at all considered temperatures.

\begin{figure}[t]
  \centering 
  \includegraphics[width=.49\textwidth]{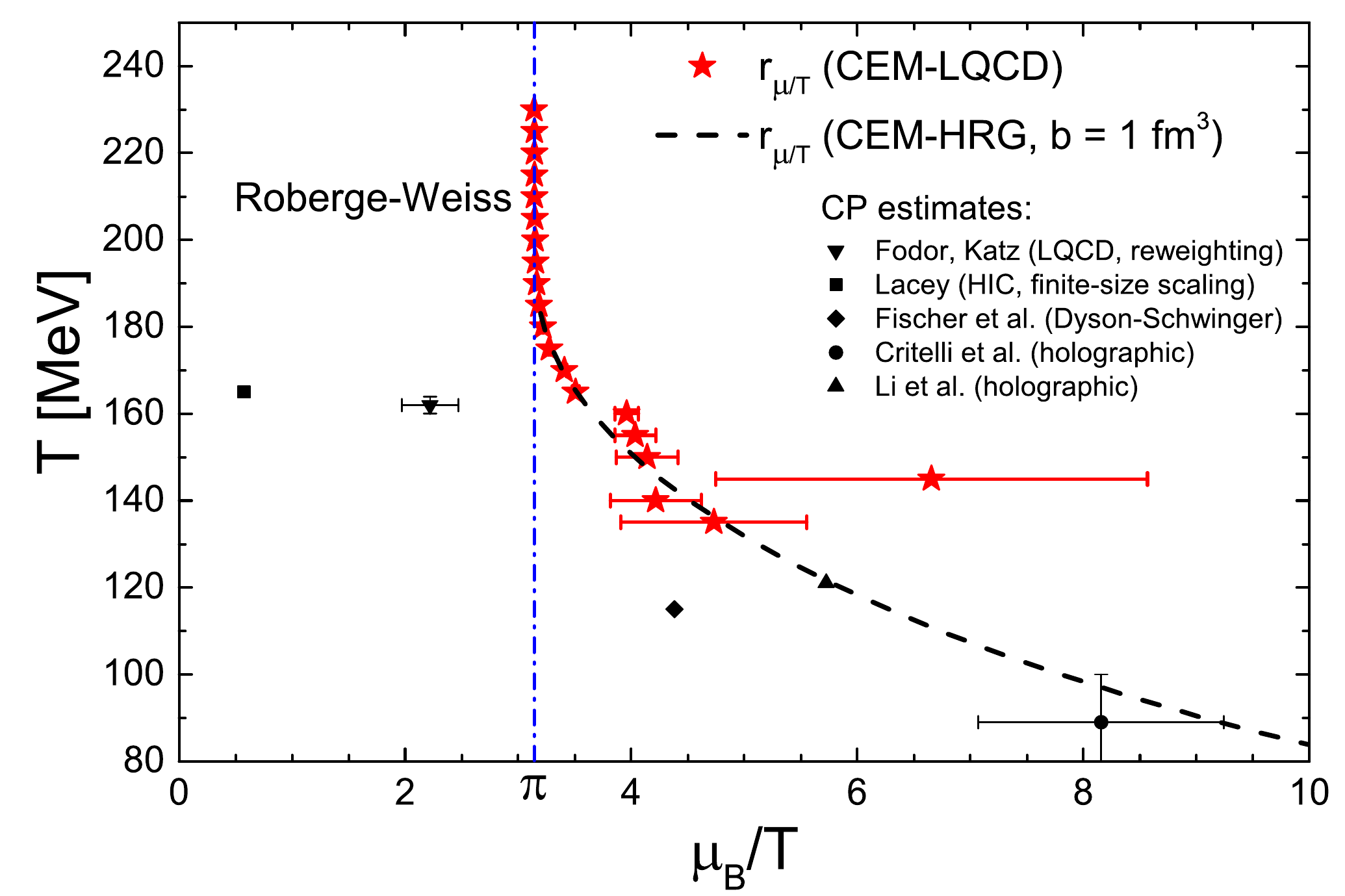}
  \caption{
  The temperature dependence of the radius of convergence $r_{\mu/T}$ of the Taylor expansion in $\mu_B/T$ of the pressure, calculated within the CEM-LQCD and CEM-HRG~($b = 1$~fm$^3$) models.
The dash-dotted blue line depicts the $\mu_B / T = \pi$ value, which corresponds to the Roberge-Weiss transition at imaginary chemical potential. 
Various QCD critical point estimates~\cite{Fodor:2004nz,Lacey:2014wqa,Fischer:2014ata,Critelli:2017oub,Li:2017ple}
are shown by the black symbols.
  } 
\label{fig:TmuTrad}
\end{figure}

The temperature dependence of the radius of convergence, $r_{\mu/T}$, 
as calculated within the CEM-LQCD~(red stars) and the CEM-HRG model~(dashed line) models using the Mercer-Roberts procedure, is presented in Fig.~\ref{fig:TmuTrad}.
$r_{\mu/T}$ is a smooth function of $T$ and it is \emph{finite}, at all temperatures considered.
The corresponding limiting singularities 
lie at complex $\mu_B / T$ values, 
as follows from
the absence of a regular asymptotic behavior of the Taylor expansion coefficients.
$r_{\mu/T}$ decreases with temperature and it approaches the asymptotic value of $r_{\mu/T} = \pi$ at higher temperatures, $T > 190$~MeV.
This value can be identified with the Roberge-Weiss~(R-W) transition~\cite{Roberge:1986mm}, which was predicted to appear at sufficiently high temperatures at imaginary chemical potential values of $\rm{Im} \, [\mu_B / T]_c = \pi (2k+1)$, and studied quite extensively in LQCD simulations~\cite{DElia:2007bkz,DElia:2009bzj,deForcrand:2010he,Wu:2013bfa,Philipsen:2014rpa,Cuteri:2015qkq,Bonati:2016pwz}.
This transition is a consequence of the R-W periodicity of the QCD partition function, $Z (\mu_B) = Z (\mu_B + i 2 \pi T )$, 
due to the center symmetry~\cite{Roberge:1986mm}, which is fully respected by the CEM.

We have cross-checked our results for $r_{\mu/T}$ by constructing Pad\'e approximants~\cite{Lombardo:2005ks,Lombardo:2006yc} for the Taylor expansion of $\chi_2^B$ in $\mu_B / T$ within the CEM-LQCD model,
and in all cases observe poles corresponding to the limiting singularity of the Taylor expansion.
These poles are located at 
$\rm{Im} [\mu_B / T]_c = \pi$, at all temperatures, while $\rm{Re} [\mu_B / T]_c$ values 
decrease towards zero at high temperatures. 
The absolute values, $|[\mu_B / T]_c|$, agree perfectly with the $r_{\mu/T}$ values in Fig.~\ref{fig:TmuTrad}.

It is interesting that numerical lattice studies at purely imaginary $\mu$ indicate $T_{\rm RW} = 208 \pm 5$~MeV for the endpoint temperature of the R-W transition~\cite{Bonati:2016pwz}, a temperature value where $r_{\mu/T}$ is already almost indistinguishable from $\pi$ in CEM-LQCD. 
We conclude that the radius of convergence of the Taylor series at $T > 135$~MeV is only determined by the singularities in the complex plane which appear to be smoothly connected to the R-W transition at high temperatures, a scenario suggested in Refs.~\cite{deForcrand:2002hgr,deForcrand:2003vyj}. 
The CEM-LQCD ``knows'' about the 
the spontaneous breaking of the center symmetry at the high temperature R-W transition indirectly, being matched to baryonic excitations
at low temperatures and to quark degrees 
of freedom 
at high temperatures. 
The exact nature and relation to the R-W transition of the singularities at intermediate temperatures still need to be clarified.
We note that CEM also inherits aspects of the chiral symmetry restoration, in the form of the input coefficients $b_1(T)$ and $b_2(T)$ taken from the lattice.

In any case, our analysis within CEM-LQCD and CEM-HRG shows no evidence for the existence of a phase transition or a critical point at real 
$\mu_B / T < r_{\mu/T}$,
with $r_{\mu/T} \geq \pi$ at all temperatures considered. 
This is consistent with all available lattice results at zero and imaginary chemical potential, 
but in contrast to various other
QCD critical point estimates available in the literature: 
these are based on lattice reweighting techniques~\cite{Fodor:2004nz}, experimental finite-size scaling analyses~\cite{Lacey:2014wqa}, the Dyson-Schwinger~\cite{Fischer:2014ata} or holographic~\cite{Critelli:2017oub,Li:2017ple} approaches, which are also shown 
in Fig.~\ref{fig:TmuTrad}. 
We note that CEM is not full QCD, therefore we do not 
rule out conclusively these other estimates.
Note also
that our results at $T < 135$~MeV are based on the HRG extrapolation of the lattice data, and therefore should be treated with care.

The particular CEM formulation presented here is simple and powerful, but it has limitations.
The relation~\eqref{eq:bk} expressing the higher-order Fourier coefficients through the first two is likely to get modified whenever effects of genuine 
many-body interactions become important. We therefore expect the model to break down at large $\mu_B / T$ values, e.g., in the dense nuclear matter region.
Note that the formalism itself can accommodate any pressure function 
periodic under the $\mu_B \to \mu_B + i \, 2\pi T$ transformation, as required by the $Z(3)$ symmetry of QCD. The CEM model can thus be extended once new and possibly contradicting lattice data become available.
However, given that CEM is consistent with all presently available lattice data we conclude that its range of applicability is at least as large as that of current lattice methods.

To summarize, a novel
Cluster Expansion Model 
for the QCD equation of state 
has been developed and applied to calculate the baryon number susceptibilities at $\mu = 0$, to very high order.
The only model inputs are the partial pressures in the $|B| = 1$ and $|B| = 2$ sectors, taken from the lattice simulations at imaginary $\mu_B$.
The model yields excellent agreement with the available lattice data for $\chi_2^B$, $\chi_4^B / \chi_2^B$, and $\chi_6^B / \chi_2^B$.
The extended model predictions for $\chi_8^B$, $\chi_{10}^B$, and $\chi_{12}^B$ shall be verified by future lattice data.
The commonly used ratio estimator is unable to determine 
the radius of convergence of the Taylor series of the pressure in $\mu_B/T$, due to a non-trivial asymptotic behavior of the Taylor coefficients.
The radius of convergence is instead determined with the more elaborate Mercer-Roberts estimator, which provides finite values of the convergence radii at all temperature values considered, $135 < T < 230$~MeV, in full agreement with the singularities of
Pad\'e approximants.
These singularities lie in the complex plane and
appear to be smoothly connected to the R-W transition at high temperatures and imaginary (baryo)chemical potential.
The analysis within CEM shows no
evidence for the existence of a phase transition or a critical point at real values of the baryochemical potential 
at $\mu_B / T \lesssim \pi$ for temperatures above 135~MeV.

The CEM model 
can be straightforwardly extended 
to calculate the equation of state of QCD at finite $\mu_B / T$, 
by supplying the $B = 0$ partial pressure $p_0(T)$ as additional model input.
Furthermore, the CEM formalism is rather flexible, and the model assumptions and input can be modified if new and contradicting lattice data becomes available.
\footnote{
After the submission of this paper it came to our attention that the series in~Eq.~\eqref{eq:chi1real} can be analytically resummed within CEM to yield
$\displaystyle \rho_B / T^3 = -\frac{2}{27\pi^2} \, \frac{\hat{b}_1^2}{\hat{b}_2} \left\{ 4 \pi^2 \, [\operatorname{Li}_1(x_+) - \operatorname{Li}_1(x_-)] + 3 \, [\operatorname{Li}_3(x_+) - \operatorname{Li}_3(x_-)] \right\}$,
where $\operatorname{Li_k}(z)$ is the polylogarithm, $\hat{b}_k \equiv b_k(T) / b_k^{\rm {_{SB}}}$, and $x_{\pm} = -\frac{\hat{b}_2}{\hat{b}_1} \, e^{\pm \mu_B / T}$.
This analytic form validates the numerical results obtained in the paper, and opens up new applications which will be explored in future works.
}


\begin{acknowledgments}

\emph{Acknowledgments.} 
Fruitful discussions with B.~Friman, C.~Greiner, V.~Koch, V.~Skokov, P.~Alba, P.M.~Lo,  and N.~Su are gratefully acknowledged.
We are grateful to B.~Friman and V.~Skokov for a suggestion to look for a resummed form of CEM.
We also thank S. Mukherjee for providing the lattice data for $\chi_2^B$, $\chi_4^B / \chi_2^B$, and $\chi_6^B / \chi_2^B$.
This work was supported by HIC for FAIR within the LOEWE program of the State of Hesse
and by the Deutsche Forschungsgemeinschaft (DFG) through the grant CRC-TR 211 ``Strong-interaction matter
under extreme conditions''
V.V. acknowledges the support from HGS-HIRe for FAIR.
H.St. acknowledges the support through the Judah M. Eisenberg Laureatus Chair at Goethe University, and the Walter Greiner Gesellschaft, Frankfurt.

\end{acknowledgments}


\end{document}